\def \SAIT #1 #2 {{\em Mem.\ Soc.\ Astron.\ It.\/} {\bf #1}, #2}
\def \MESS #1 #2 {{\em The Messenger\/} {\bf #1}, #2}
\def \ASTRNACH #1 #2 {{\em Astron. Nach.\/} {\bf #1}, #2}
\def \AAP #1 #2 {{\em Astron. Astrophys.\/} {\bf #1}, #2}
\def \AAL #1 #2 {{\em Astron. Astrophys. Lett.\/} {\bf #1}, L#2}
\def \AAR #1 #2 {{\em Astron. Astrophys. Rev.\/} {\bf #1}, #2}
\def \AAS #1 #2 {{\em Astron. Astrophys. Suppl. Ser.\/} {\bf #1}, #2}
\def \AJ #1 #2 {{\em Astron. J.\/} {\bf #1}, #2}
\def \ANNREV #1 #2 {{\em Ann. Rev. Astron. Astrophys.\/} {\bf #1}, #2}
\def \APJ #1 #2 {{\em Astrophys. J.\/} {\bf #1}, #2}
\def \APJL #1 #2 {{\em Astrophys. J. Lett.\/} {\bf #1}, L#2}
\def \APJS #1 #2 {{\em Astrophys. J. Suppl.\/} {\bf #1}, #2}
\def \APSS #1 #2 {{\em Astrophys. Space Sci.\/} {\bf #1}, #2}
\def \ASR #1 #2 {{\em Adv. Space Res.\/} {\bf #1}, #2}
\def \BAIC #1 #2 {{\em Bull. Astron. Inst. Czechosl.\/} {\bf #1}, #2}
\def \JSQRT #1 #2 {{\em J. Quant. Spectrosc. Radiat. Transfer\/} {\bf #1}, #2}
\def \MN #1 #2 {{\em Mon. Not. R. Astr. Soc.\/} {\bf #1}, #2}
\def \MEM #1 #2 {{\em Mem. R. Astr. Soc.\/} {\bf #1}, #2}
\def \PLR #1 #2 {{\em Phys. Lett. Rev.\/} {\bf #1}, #2}
\def \PASJ #1 #2 {{\em Publ. Astron. Soc. Japan\/} {\bf #1}, #2}
\def \PASP #1 #2 {{\em Publ. Astr. Soc. Pacific\/} {\bf #1}, #2}
\def \NAT #1 #2 {{\em Nature\/} {\bf #1}, #2}

\documentstyle{memsait}
\input epsf.sty
\begin{opening}
\title{PAIR CREATION AT SHOCKS: APPLICATION TO AGNs} 
\author{P.O. PETRUCCI$^1$, G. HENRI$^2$, G. PELLETIER$^2$}
\institute{$^1$ OAB, Milano, Italy\\$^2$ LAOG, Grenoble, FRANCE}
\date{} 
\end{opening}

\begin{document}

\oddpagefooter{}{}{} 
\evenpagefooter{}{}{} 

\begin{abstract}
We investigate the effect of pair creation on a shock
structure. Actually, particles accelerated by a shock can be sufficiently
energetic to boost, via Inverse Compton (IC) process for example,
surrounding soft photons above the rest mass electron energy and thus to
trigger the pair creation process. The increase of the associated pair
pressure is thus able to disrupt the plasma flow and possibly, for too
high pressure, to smooth it completely. Reversely, significant changes of
the flow velocity profile may modify the distribution function of the
accelerated particles in such a way that the number of particles enable
to trigger the pair creation process may change, modifying consequently
the pair creation rate. We propose here, using simplifying assumptions,
to study the global behavior of a shock structure in an environment
dominated by pairs.
\end{abstract}

\section{Geometry and main hypotheses}
The schematic view of our toy model can be seen in Fig. 1.
\begin{figure}[h]
\hspace*{0cm}
\begin{tabular}{ll}
\begin{minipage}{8cm}
\epsfxsize=7cm
\hspace{0cm}\epsfbox{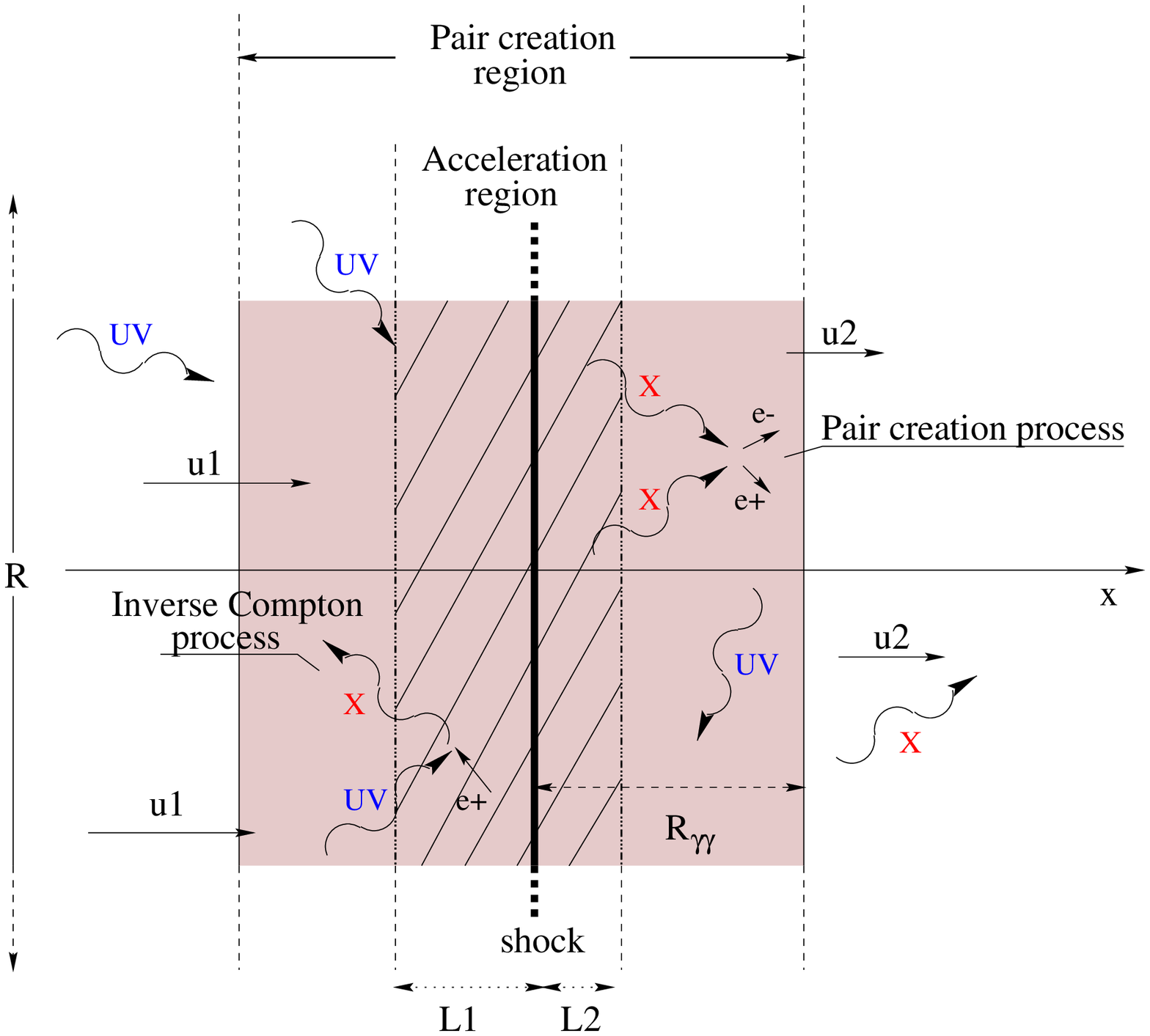}
\end{minipage} &
\hspace*{-1cm}
\begin{minipage}{5cm}
\caption[h]{Schematic view of a shock. The shock discontinuity is
represented by the vertical bold line.  We have also indicated the
different parameters defining the acceleration and pair creation
region. Particles are represented by straight arrows and photons by
warped ones. Scales are not respected. \label{mainchoc}}
\end{minipage}
\end{tabular}
\vspace*{-0.5cm}
\end{figure}
We assume a 1D geometry and we suppose the shock to be located in
$x=0$. This will be insured by imposing that the flow velocity possesses
an inflection point in $x=0$ that is (noting $u(x)$ the flow velocity):
$\displaystyle\left.\frac{\partial^2 u}{\partial x^2}\right|_{x=0}=0
\ (1)$.  We suppose the spatial diffusion coefficient $D$ to be independent
of $x$ and of the energy of the particles and we assume the plasma
pressure $P_{tot}$ to be dominated by the pressure $P_{rel}$ of the
relativistic leptons population.

\section{Shock disappearance}
The shock still exists as long as Eq. (1) (which depends on the pair
pressure creation rate $\dot{P}_{\pm}$ and on the size of the pair
creation region $R_{\gamma\gamma}$ cf. Fig. 1) is verified. It appears
that, for a given value of $R_{\gamma\gamma}$, there exists a limit value
$\dot{P}_{\pm}^{lim}$ of $\dot{P}_{\pm}$ above which Eq. (1) has no
consistent solutions anymore and the shock disappears.

\section{Variability}

As seen before, for pair pressure creation rate larger than
$\dot{P}_{\pm}^{lim}$ the shock disappears. However, with the
corresponding fall down of the acceleration processes, the rate of pair
creation, and consequently the pair pressure creation rate, decreases and
the shock may appear again, initiating pair processes until a new
destruction of the shock. These cycles shock appearance/disappearance
could thus be the origin of the high energy variability observed in AGNs
(and perhaps in prompt $\gamma$-ray bursts where the compactness may be
large enough).

\section{Stationary states}
If the pair pressure modifies the flow velocity profile, reversely a
 change of the flow velocity profile can modify the distribution function
 of the accelerated particles in such a way that the number of particles
 enable to trigger the pair creation process (i.e.  particles with
 $\gamma\ge\gamma_{th}$) may change, modifying consequently the pair
 creation rate. Stationary states are then obtained by solving
 self-consistently for the particle distribution function $n(\gamma)$ and
 the flow velocity profile $u(x)$, linked through the pair pressure
 creation rate. We have reported on Fig. 2 the contour plots of the
 spectral index and high energy cut-off of the high energy spectra
 produced by IC for stationary states characterized by different values
 of the soft compactness and the upstream flow velocity.
\begin{figure}[t]
\epsfxsize=6cm
  \begin{tabular}{ll}
  \epsfbox{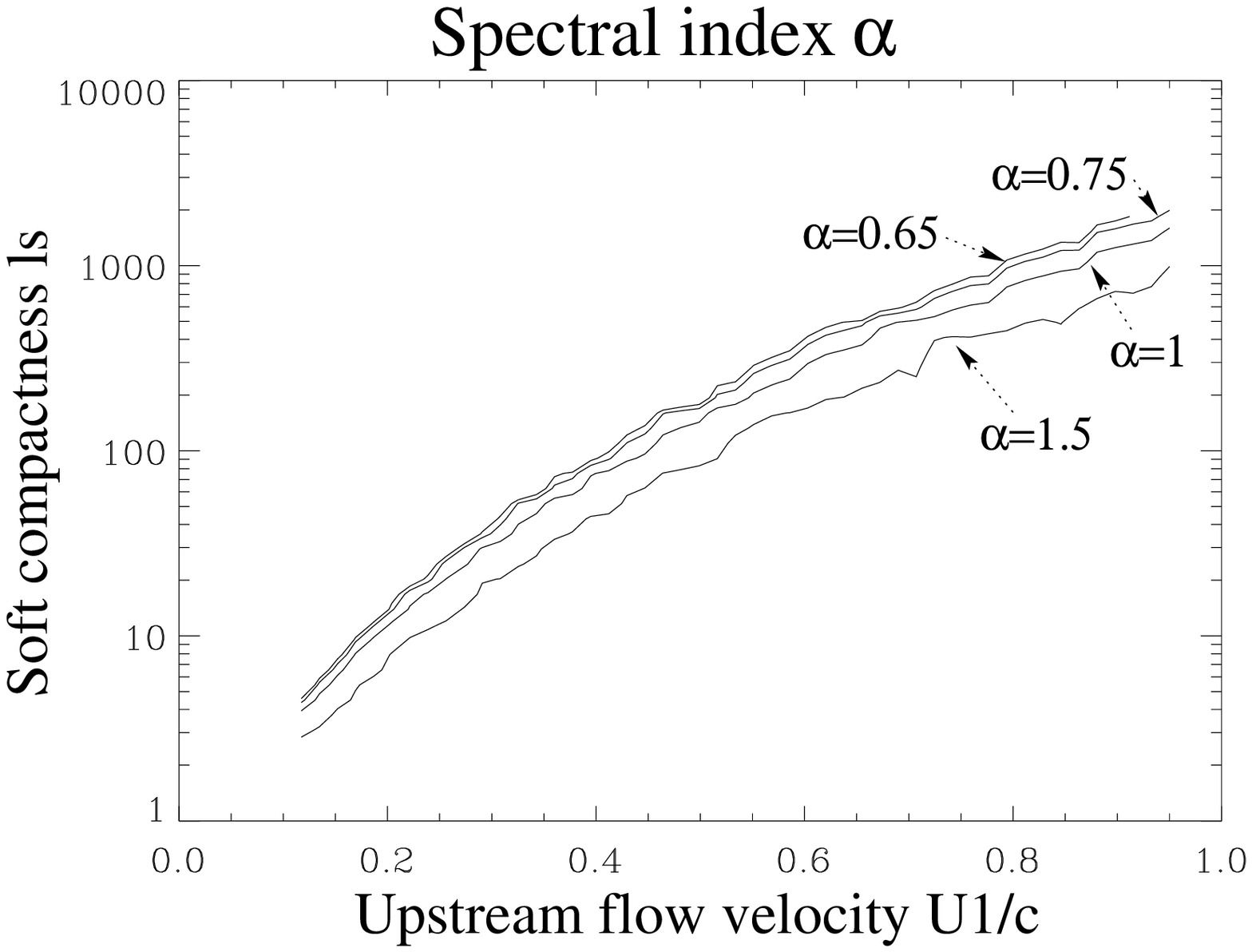}&\epsfbox{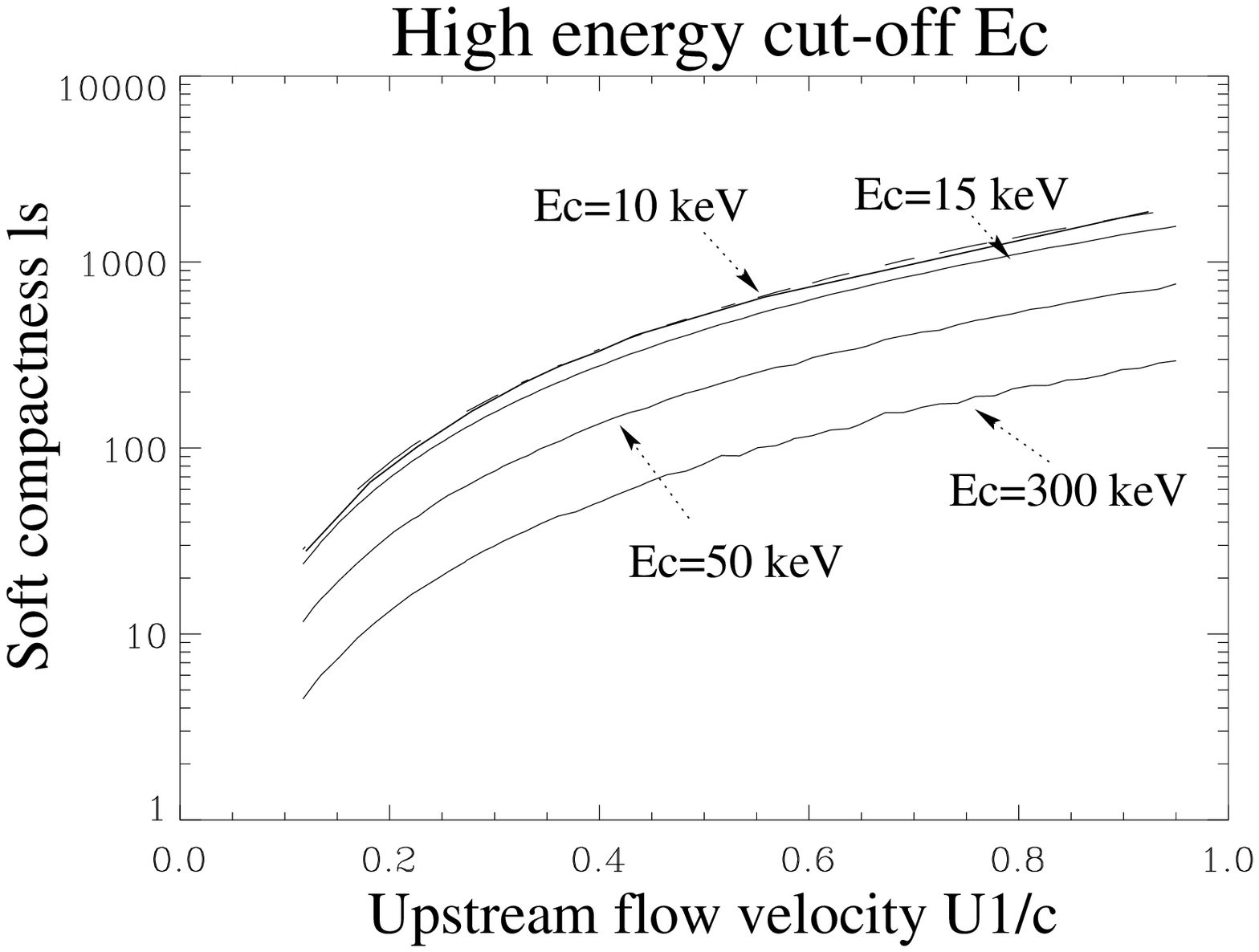}\\
  \end{tabular}
\caption[h]{Contour plots in the ($l_s$, $u_1/c$) space of the spectral
  index $\alpha$ and the high energy cut-off $E_c$ of the high energy
spectrum ($ F_E\propto E^{-\alpha}\exp\left(-\frac{E}{E_c}\right)$)
emitted by IC by the accelerated particles when th system is in
stationary state. We assume a soft photon energy of 1 eV}
\vspace{-0.5cm}
\end{figure}

\end{document}